\documentclass[12pt,preprint]{aastex}
\slugcomment{Version V4.1 2-Jan-2010}%

\shorttitle{The {\it Kepler} Follow-up Program}
\shortauthors{T. N. Gautier et al.}

\begin{document}

\title{The {\it Kepler} Follow-up Observation Program}

\author{Thomas~N.~Gautier~III\altaffilmark{1},
Natalie~M.~Batalha\altaffilmark{2},
William~J.~Borucki\altaffilmark{3},
William~D.~Cochran\altaffilmark{4},
Edward~W.~Dunham\altaffilmark{5},
Steve~B.~Howell\altaffilmark{6},
David~G.~Koch\altaffilmark{3},            %
David~W.~Latham\altaffilmark{7},
Geoff~W.~Marcy\altaffilmark{8},
Lars~A.~Buchhave\altaffilmark{7,9},
David~R.~Ciardi\altaffilmark{10},
Michael~Endl\altaffilmark{4},
Gabor~F\H{u}r\'{e}sz\altaffilmark{7},
Howard~Isaacson\altaffilmark{8},
Phillip~MacQueen\altaffilmark{4},
Georgi~Mandushev\altaffilmark{5},
Lucianne~Walkowicz\altaffilmark{8} }

\altaffiltext{1}{Jet Propulsion Laboratory/California 
Institute of Technology, 4800 Oak Grove Dr, Pasadena, California 91109; Thomas.N.Gautier@jpl.nasa.gov}
\altaffiltext{2}{Department of Physics and Astronomy, San Jose State University, San Jose, CA 95192}
\altaffiltext{3}{NASA Ames Research Center, Moffett Field, CA 94035}
\altaffiltext{4}{McDonald Observatory, The University of Texas at Austin, Austin, TX 78712}
\altaffiltext{5}{Lowell Observatory, Flagstaff, AZ 86001}
\altaffiltext{6}{National Optical Astronomy Observatory, Tucson, AZ 85719}
\altaffiltext{7}{Harvard-Smithsonian Center for Astrophysics, Cambridge, MA 02138}
\altaffiltext{8}{University of California, Berkeley, Berkeley, CA 94720}
\altaffiltext{9}{Niels Bohr Institute, Copenhagen University, DK-2100 Copenhagen, Denmark}
\altaffiltext{10}{NASA Exoplanet Science Institute, California Institute of Technology, Pasadena, CA 91125}

\begin{abstract}

     The {\it Kepler Mission} was launched on March 6, 2009 to perform a photometric 
survey of more than 100,000 dwarf stars to search for terrestrial-size planets 
with the transit technique. Follow-up observations of planetary candidates 
identified by detection of transit-like events are needed both for 
identification of astrophysical phenomena that mimic planetary transits 
and for characterization of the true planets and planetary systems found 
by {\it Kepler}. We have developed techniques and protocols for detection of false 
planetary transits and are currently conducting observations on 177 {\it Kepler} targets that 
have been selected for follow-up. A preliminary estimate indicates that between 
24\% and 62\% of planetary candidates selected for follow-up will turn out to 
be true planets.

\end{abstract}


\section{Introduction}\label{intro}

     {\it Kepler}, a NASA Discovery Mission designed to determine the frequency of 
terrestrial-size planets in and near the habitable zone of 
solar-like stars, was launched on March 6, 2009 and has been returning scientific data 
since early May 2009. The specific goals of the mission are:
\begin{itemize}
\item Determine the frequency of Earth-size and larger planets in 
or near the habitable zone of a wide variety of stars.
\item Determine the distributions of planet sizes and orbital semi-major axes
\item Estimate the frequency and orbital distribution of planets in 
multiple-star systems.
\item Determine the distributions of semi-major axis, albedo, size, 
mass, and density of short-period planets giant planets.
\item Identify additional members of each photometrically discovered 
planetary system using complementary techniques.
\item Determine the association of stellar properties with planetary characteristics. 
\end{itemize}

{\it Kepler} will survey more than 100,000 late-type dwarf stars in the solar 
neighborhood with visual magnitudes between 9 and 16 for a period of 3.5 
years looking for transits of planets around those stars.  
Details of the {\it Kepler Mission}, the photometer and its operating modes 
are given in \citet{bor10} and \citet{mission}.

While the {\it Kepler} photometer is capable of detecting transits of 
Earth-size planets around its target stars the photometric detections 
must be supplemented by follow-up observations with other facilities 
to identify transit-like signals from non-planetary sources and to 
accomplish goals of the mission beyond just detection of the planets. 
The {\it Kepler} Follow-up Observation Program (FOP) is designed to provide 
these supplemental observations.

\section{Purpose of {\it Kepler} Follow-up Program}\label{purpose}

     The detection of a transit-like signal in a {\it Kepler} target star is not 
sufficient evidence to confirm the presence of a planet orbiting the 
target star, nor would it be for any transiting planet survey. \citet{brown03} 
distinguishes 12 combinations of giant planets and stars in eclipsing 
and transiting systems that can produce light curves mimicking a planet 
transiting a solitary primary star. Six of the combinations do not involve 
planets at all and four others distort the transit light curve so that the 
size of the planet is indeterminate. In the case of {\it Kepler} where it is 
practical to detect transits of planets down to the size of Earth and 
below additional opportunities for confusion exist.

The principal causes of false positive or ambiguous planet size detections 
for {\it Kepler} are:

\begin{enumerate}
\item Background eclipsing binaries where a distant eclipsing star system's light 
is confused with and diluted by the light of the {\it Kepler} target star. The deep 
eclipse of the binary will appear as a shallow transit characteristic of a 
planet around the target star.

\item Eclipsing binaries in multiple-star systems where the multiple system is 
the {\it Kepler} target but the deep eclipse of the binary is diluted by the light 
of other stars in the system so as to appear as shallow as a planetary transit.

\item Grazing eclipses of binary stars that are {\it Kepler} target stars

\item Transits in a binary system consisting of a giant star and a main sequence star

\item Transits of background main sequence stars by giant planets that are diluted by the 
light of the {\it Kepler} target star so as to appear as shallow as transits of Earth-size planets

\item Transits of giant stars by giant planets that appear as shallow as Earth-size 
planets transiting main sequence stars.
\end{enumerate}

In general light curve data alone cannot distinguish between these six 
configurations and the principle objective of the {\it Kepler} survey, Earth-size planets 
transiting main sequence stars. {\it Kepler} data is collected in a way that allows 
determination of the photo-center of light around the target star which can often identify 
causes 1 and 5 above (see sections \ref{K-data} and \ref{proto} below). Careful spectroscopic 
and photometric follow-up observations of candidate transiting planets are needed to fully 
determine the source of a {\it Kepler} transit signal.

The {\it Kepler} follow-up observation program therefore has two purposes. First, the 
follow-up program must separate false positive transit indications from true 
planets to a high degree of reliability. Second, the follow-up program will 
characterize a representative sample of planets and planetary systems by 
determining the mass and orbit of the transiting planet and any other detectable, 
non-transiting planets.

\section{False Positive Elimination}\label{fpelim}

\subsection{Elimination with {\it Kepler} data}\label{K-data}

     False positive elimination for {\it Kepler} transit detections begins in the 
{\it Kepler} Pipeline \citep{pipeline} using only {\it Kepler} data. After 
transit-like signals are detected and matched against each other to find 
periodic series of transits, the series, now called Threshold Crossing 
Events (TCEs), are checked for adequate signal to noise ratio (SNR), regularity 
of period, transit depth, transit shape and consistency with Kepler's 
(the astronomer) laws of planetary motion. TCEs that pass these tests are 
subject to a preliminary photocenter motion analysis wherein the motion 
of the centroid of the light distribution in the photometer aperture, in 
and out of transit, is measured. Large 
centroid motions likely indicate that a nearby object, not the {\it Kepler} target, 
is the source of the transit signal. In these cases examination of the light curves 
from individual pixels in the photometer aperture usually identifies the background 
object. These steps will attack false positive causes 1 and 5 in section \ref{purpose}.

     The full light curve of the target star, in and out of transit, is examined by a 
human to look for characteristic signs of an eclipsing binary star or other 
behavior unexpected of a transiting planet. A power spectrum of the light 
curve may be examined to determine if the target star is a giant or a dwarf as giants will 
display noise levels typically an order of magnitude higher than dwarfs 
\citep[see, for example, the analysis of KOI-145 in][]{seismo}.
(The {\it Kepler} planetary target set is intended to exclude giants except in specific 
cases; see \citet{targets}. However, some giants will inevitably have leaked in.) 
Examination of the light curve can also help detect grazing transits of stellar companions 
or giant planets which can be mistaken for transits of small planets. These steps will attack 
causes 2, 3 and 4 in section \ref{purpose}.

     All of these pre-FOP steps are described more fully in the companion paper by \citet{tcert}. 
At the time of this writing these steps have, for the most part, been carried 
out manually. With the delivery of the next version of the {\it Kepler} Pipeline 
many of the steps will be automated.
     
\subsection{Elimination with Follow-up Observations}

Planetary candidates that survive elimination on the basis of the {\it Kepler} 
data are sent for follow-up observation (see section \ref{ops} below) where 
moderate precision ($\sim 300$ m/s), moderate SNR radial velocity (RV) 
measurements are used to detect and eliminate stellar mass companions as the cause 
of the transit signal. Spectra for these measurements 
are usually taken at high resolution, as for high precision radial velocity work, but 
without enhancements such as exposure through an iodine cell that enable high RV precision. 
Imaging under good seeing conditions, active optic imaging and speckle imaging 
are used in combination with the previously obtained centroid motion 
analysis to determine if the {\it Kepler} target is the true source of the 
transit signal. Moderate precision, high SNR spectra may be used for 
line bisector analysis to detect multiple-star systems 
\citep[see][and references therein]{tor04}. 
The moderate precision spectra are also used for classification 
of the target stars. These methods are generally the same as those discussed by \citet{alon04}. 

These follow-up observations will detect false positives due to eclipsing 
binary stars both in the background of the {\it Kepler} target and as the target 
star itself, either as a solitary binary or in a multiple-star system where 
a third light dilutes the transit. Planetary candidates surviving all of 
these observations and tests should have a high probability of being a true 
planet.

The current gold standard for extrasolar transiting planet confirmation is radial 
velocity determination of an orbit for the planet that is consistent with the 
timing and duration of the observed transits. Measurement of the planet's mass 
is also confirmatory. This sort of determination will be possible only for a fraction 
of the planets anticipated to be detected by {\it Kepler}, given the limitations of 
the sensitivity of the RV method and the resources available to {\it Kepler}. 
Giant planets around late type main sequence stars with orbital periods up to a few 
years and Earth mass planets in short period orbits around low mass stars will produce 
enough reflex motion in their parent stars to be measurable with current spectroscopic techniques. 
Some examples of this kind of confirmation for {\it Kepler} planet detections can be 
found in papers in this volume \citep{kep4, kep5, kep6, kep7, kep8}. The low 
mass planets in long period orbits expected to be detected by {\it Kepler} will not 
produce enough reflex motion to be seen by current instrumentation. Additionally, 
the bulk of {\it Kepler} planetary survey targets are fainter than about 13.5 visual 
magnitude making high precision radial velocity measurements prohibitively expensive 
or impossible. Confidence in most low mass planets found by {\it Kepler} will 
have to be based on the quality of the photometric data and elimination of false positive 
possibilities.

\section{Operation of Follow-up Program}\label{ops}

     As the results of the analyses of the {\it Kepler} data described 
in section 3.1 become available, the {\it Kepler} TCE Review Team (TCERT) 
convenes to review the results and select TCEs that are likely to be true 
transiting planets. The selected TCEs are assigned {\it Kepler} Object of Interest 
numbers (KOIs) and the TCERT requests the FOP Observer Group (FOG) to begin 
false positive elimination on the KOIs. These requests 
include a statement of priority and a discussion of the 
analytical results that led to the selection of the TCEs and KOIs.

     When requests are received from the TCERT the FOG reviews 
the priorities and analyses and begins observations of 
the KOIs following a protocol for false positive elimination. As data and 
results of the follow-up observations accumulate, the FOG discusses 
the status of each KOI under active observation to monitor the course of 
the observations. When work is completed and the FOG has decided the disposition 
of the KOI the results are passed back to the TCERT and the {\it Kepler} Science 
Analysis System (KSAS, described below).

     The TCERT then reviews the results of false positive elimination and 
may ask the FOG to perform more observations on selected KOIs to further 
characterize a planet or planetary system. 

\subsection{Follow-up Protocol}\label{proto}

   The protocol for elimination of false positives consists of a 
series of observations with 1-3 meter class telescopes with good 
spectroscopic or high spatial resolution capability.

\begin{enumerate}
\item Begin with a reconnaissance spectrum taken at moderate precision 
and with moderate signal to noise ratio to provide radial velocity 
precision of $\sim300$ m/s. This spectrum will verify and improve 
stellar parameters from the {\it Kepler Input Catalog} and 
determine the star's rotation velocity. Fast rotating stars 
($v\sin{i} \geq 15$ km/s) and hot stars, early F and earlier, that show 
nearly featureless spectra are more difficult for high precision RV 
measurements and may be given low priority for characterization studies. 
Double-lined spectroscopic binaries that escaped pre-FOP detection due to 
grazing eclipses shallow enough to be mistaken for planetary transits may 
be detected at this point. This reconnaissance spectrum is also the first 
point in the RV time series needed to detect stellar mass companions. 
This step attacks false positive causes 3 and 6 described in section \ref{purpose}.

\item {Obtain images of the field around the KOI using active optics, 
speckle imaging and conventional imaging to search for background objects 
that might be the source of the transit signal. Out-of-transit images provide the light 
distribution in the {\it Kepler} photometric aperture which allows 
detailed interpretation of the centroid motion analysis described in 
section \ref{K-data}. This analysis can often distinguish if the KOI is the 
source of the signal. \citet{kep7}, \citet{kep6}, \citet{kep5} and \citet{kep8} 
give examples of this analysis.

    In-transit images, though harder to obtain due to observation scheduling 
difficulties, will occasionally be useful. 
For large transit signals, $\sim 1$\%, in-transit photometry can provide 
a definitive indication of which object in the aperture is producing the 
transit signal by observing which object dims during transit. For small signals, 
$\sim0.01$\% as expected for Earth-size 
planets, where centroid motion analysis may be too insensitive for definite 
determination of the signal source, in-transit photometry can provide 
falsification of background objects as the signal source by showing that any 
observed background object variations are too small to provide the transit 
signal \citep[after][]{corot}.}

This step attacks causes 1 and 5 in section \ref{purpose}.

\item If the KOI is still of interest, continue moderate precision RV 
measurements. Stellar mass companions responsible for the transit signal 
will reveal themselves by large velocity variations consistent with an orbit 
commensurate with the transit light curve. This step attacks causes 3 and 4 in 
section \ref{purpose}.

\item Obtain a short time series of high SNR, moderate resolution spectra 
for line bisector analysis to detect KOIs which are multiple-stars where 
the transit signal is diluted by non-transited components of the system. 
This step attacks cause 2 in section \ref{purpose}.
\end{enumerate}

  This protocol also detects dilution of transits of true giant 
planets and can allow for correction of the dilution for proper 
measurement of planetary diameter.

    KOIs determined to be background eclipsing binary stars will generally be eliminated 
from further follow-up observation. KOIs in which the target star is found to be an 
eclipsing binary may be removed from further observations if RV techniques cannot yield 
a planet confirmation. However, eclipsing binary systems are inherently interesting for a 
transiting planet search since the planetary plane is likely to be close to the binary plane. The 
TCERT may ask members of the {\it Kepler} Science Team to use other methods, such as 
transit timing variation, to search for planets around these stars.
    
\subsection{Further characterization}

    After the FOP has reported to the TCERT that a KOI is likely 
to be a planet the TCERT may request, as mentioned above, that further 
characterization of the KOI be done. These requests are expected to be 
mainly for determination of planetary orbits and masses or for 
precise measurements of stellar and planetary parameters, including 
planetary atmosphere temperature and composition. RV searches for additional, 
non-transiting planets may also be made. Methods employed include:

\begin{itemize}
\item High precision RV measurements (1-10 m/s) for orbit and mass 
determination and for searches for non-transiting planets. 
Measurements of the Rossiter-McLaughlin effect \citep{ros24, mcl24, kop90}
may be made, especially for stars with higher rotation rates 
for which RV is more difficult and RM is more sensitive. We have currently obtained 
RM data for one planet, Kepler-8b \citep{kep8}.

\item Observation of secondary eclipses with the Spitzer Space Telescope 
to measure temperatures in the planetary atmosphere and to determine 
atmospheric composition.

\end{itemize}

\subsection{Planet Confirmation}

    The {\it Kepler} team is currently working to develop and validate methods 
other than spectroscopic orbit determination for confirmation of transiting 
planets. Orbit determination is excessively 
expensive or impossible for the majority of planets expected to be detected 
by {\it Kepler}. Therefore, analysis of {\it Kepler} photometry combined with follow-up 
observations using moderate precision spectroscopy and high spatial resolution imaging, 
as described in section \ref{proto}, or other methods that may be developed must be used 
alone to eliminate false positives, leaving 
a statistically reliable, but not perfect, set of planet detections. The standard method of 
spectroscopic orbit determination and Rossiter-McLaughlin detection 
are being used on the brighter {\it Kepler} targets harboring giant planets and 
low mass planets in short period orbits around low mass stars to determine 
the reliability of planets surviving the protocol of section \ref{proto}.

   We expect that {\it Kepler'}s extraordinary photometric precision, which allows detection 
of the odd-even effect in binary star eclipses, transit shapes that are inconsistent with 
planets and secondary occultations of actual transiting planets, can be combined with centroid 
motion analysis and follow-up spectroscopy to achieve the required degree of false positive 
rejection.

\subsection{Resources}

    The follow-up program has a team of 16 observers and 5 other 
astronomers who dedicate at least part of their time to {\it Kepler} follow-up 
observations. The instrument facilities available are shown in Table 
\ref{instr_table}. In the 2009 observing season from June through 
November 13 nights of Keck time and more than 100 nights on other 
telescopes were used for {\it Kepler} follow-up observations.

\section{Current Results}\label{results}

     At the end of the 2009 observing season 177 Kepler targets 
have been identified as KOIs from commissioning (Q0) and quarter 1 (Q1) 
data sets and sent to the FOP for follow-up observation 
\citep[see][for a description of these data]{tcert,targets}. 
These 177 KOIs were selected from the 
Kepler planetary targets brighter 
than ${\rm 15^{th}}$ Kepler magnitude and have been classified as shown in Table \ref{KOI_table}. 
This table is intended only to present the current status of Kepler follow-up 
observation and is not suitable for drawing statistical inferences. 
Obviously interesting numbers like completeness estimates for planet 
detection cannot yet be sensibly derived. Final versions, optimized for actual flight data, 
of the algorithms for transit detection, background eclipsing binary elimination and other 
procedures described in section \ref{K-data} were not in place when these KOIs were selected. 
Indeed some steps of the selection were performed by human examination of the light 
curves and the number of background eclipsing binaries reflects early, inefficient 
versions of pre-FOP elimination. When improved versions of the algorithms are applied 
to the Q0 and Q1 data new KOIs may appear which will be followed up in future observing 
seasons. Sensible preliminary estimates of survey completeness 
and other statistics should begin to be available after the end of the 2010 observing season when 
our KOI selection methods are better understood and more KOIs currently under reconnaissance 
are resolved.

     A preliminary estimate of the fraction of planets expected 
in the KOIs sent for follow-up observation is possible. A well studied set of 
70 transit detections from planetary targets brighter than ${\rm 14^{th}}$ magnitude have 
been subject to our current best versions of detection validation and background 
eclipsing binary elimination as described in section \ref{K-data}. Table 
\ref{stat_table} presents the 
state of follow-up observations for these 70 targets. Seventy percent of the 
initially interesting transit detections were found to be false positive in the 
pre-FOP vetting process. Of the 21 KOIs left for FOP follow-up, at least 24\% are 
good planets. All 8 of the KOIs not yet rejected nor confirmed as planets could 
prove to be good planets, providing a first estimate of between 24 and 60\% good 
planets in KOIs sent for follow-up observation. Small number statistics and incomplete 
analysis of the 8 KOIs currently prevent a better estimate.

\acknowledgements

    We gratefully acknowledge the outstanding work of the enormous number of people 
on the {\it Kepler} team who have contributed to the success of the mission, and 
without whom the Follow-up Program would have nothing to follow.

     {\it Kepler} was competitively selected as the tenth Discovery mission. Funding for 
this mission is provided by NASA's Science Mission Directorate. The NASA Exoplanet 
Science Institute developed the FOP coordination data server with the capable services of Megan Crane 
as developer and David Imel as manager.

\noindent{\it Facilities:} \facility{The Kepler Mission}

\clearpage

\begin{deluxetable}{llll}
\tabletypesize{\scriptsize}
\tablecaption{Follow-up Resources \label{instr_table}}
\tablewidth{0pt}
\tablehead{
\colhead{Telescope} & \colhead{Instrument}& \colhead{Institution}& \colhead{Location}
}
\startdata
Tillinghast reflector     & TRES Spectrograph            & CfA                   & Arizona\\
Shane telescope           & Hamilton Spectrograph        & Lick observatory      & California\\
Nordic Optical Telescope  & FIES Spectrograph            &                       & Canary Islands\\
Keck Telescope            & HIRES Spectrograph           & NASA                  & Hawaii\\
Palomar 5m Hale Telescope & PHARO/NGS AO camera          & Caltech/JPL           & California\\
WIYN Telescope            & Speckle Camera               & KPNO                  & Arizona\\
KPNO 2.1m Telescope       & Gcam Spectrograph            & KPNO                  & Arizona\\
KPNO 4m Telescope         & RCSpec                       & KPNO                  & Arizona \\
Hobby Eberly Telescope    & High Resolution Spectrograph &                       & Texas\\
Harlan J. Smith Telescope & Tull Coude Spectrograph      & McDonald Observatory  & Texas\\
MMT                       & ARIES AO system              & SAO/Univ. Arizona     & Arizona\\
1.1-m Hall                & NASA42 CCD Camera            & Lowell Observatory    & Arizona\\
1.8-m Perkins             & PRISM CCD Camera             & Lowell Observatory    & Arizona\\
William Herschel Telescope& HARPS NEF spectrometer (when completed) &            & Canary Islands\\
Spitzer Space Telescope   & IRAC camera (warm)           & NASA                  & Earth trailing orbit\\
Hubble Space Telescope    &                              & NASA                  & low Earth orbit\\
\enddata
\end{deluxetable}

\begin{deluxetable}{lrl}
\tabletypesize{\scriptsize}
\tablecaption{Current Status of Observations \label{KOI_table}}
\tablewidth{7in}
\tablehead{ \colhead{Type} & \colhead{Number}& \colhead{} }
\startdata

Total KOIs & 177 & From targets $m_{kepler}\leq15$ in quarters 0 and 1\\
 & & \\
Planet                  &  5 & Good rv orbit matches light curve.\\
Possible planet         & 52 & Radial velocity variation is small enough for a planetary mass companion.\\
Recon                   & 65 & Still under reconnaissance. No type assigned.\\
Double lined spectrum   &  5 & \\
Stellar companion       &  8 & RV variations indicate a stellar mass companion.\\
Triple system           &  1 & Transit source is in a triple (or greater) system.\\
Background eclipsing binary & 11 & \\
Fast rotator            & 13 & Star is rotating too fast for very precise velocities.\\
Withdrawn               & 14 & Withdrawn by TCERT after re-examination of light curve\\
Unsuitable              &  3 & Featureless spectrum unsuitable for RV work or no star apparent at target location\\
\enddata
\end{deluxetable}

\begin{deluxetable}{lrr}
\tabletypesize{\scriptsize}
\tablecaption{False Positive Rejection Statistics \label{stat_table}}
\tablewidth{7in}
\tablehead{ \colhead{Type} & \colhead{Number}& \colhead{Fraction} }
\startdata
Number of targets in well studied sample (see text)   & 70 &      \\
Rejected by photometric appearance or centroid motion & 49 & 70\% \\
 & & \\
KOIs left after pre-follow-up vetting                 & 21 &      \\
Planet                                                &  5 & 24\% \\
Not yet rejected nor confirmed as planets             &  8 & 38\% \\
Rejected by FOP observation                           &  7 & 33\% \\
Dropped due to confusion with nearby stars            &  1 &  5\% \\
\enddata
\end{deluxetable}

\end{document}